# Three-dimensional band structure of layered TiTe$_2$: Photoemission final-state effects

V.N. Strocov,[1,*] E.E. Krasovskii,[2] W. Schattke,[2,3] N. Barrett,[4] H. Berger,[5] D. Schrupp[6] and R. Claessen[6,7]

[1] *Swiss Light Source, Paul Scherrer Institute, CH-5232 Villigen PSI, Switzerland*
[2] *Institut für Theoretische Physik, Christian-Albrechts-Universität, D-24098 Kiel, Germany*
[3] *Donostia International Physics Center, 20018 San Sebastian, Basque Country, Spain*
[4] *CEA-DSM/DRECAM-SPCSI, CEA-Saclay, 91191 Gif-sur-Yvette, France*
[5] *Institut de Physique de la Matière Complexe, EPFL, CH-1015 Lausanne, Switzerland*
[6] *Experimentalphysik II, Universität Augsburg, D-86135 Augsburg, Germany*
[7] *Experimentelle Physik 4, Universität Würzburg, D-97074 Würzburg, Germany*

Three-dimensional band structure of unoccupied and occupied states of the prototype layered material TiTe$_2$ is determined focusing on the ΓA line of the Brillouin zone. Dispersions and lifetimes of the unoccupied states, acting as the final states in the photoemission process, are determined from a very-low-energy electron diffraction experiment supported by first-principles calculations based on a Bloch waves treatment of multiple scattering. The experimental unoccupied states of TiTe$_2$ feature dramatic non-free-electron effects such as multiband composition and non-parabolic dispersions. The valence band layer-perpendicular dispersions are then determined from a photoemission experiment consistently interpreted on the basis of the experimental final states to achieve control over the 3-dimensional wavevector. The experimental results demonstrate the absence of the Te 4$p_z$* Fermi surface pocket at the Γ point and significant self-energy renormalization of the valence band dispersions. Photoemission calculations based on a novel Bloch waves formalism within the one-step theory reveal limitations of understanding photoemission from layered materials such as TiTe$_2$ in terms of direct transitions.

PACS numbers: 79.60.-i, 71.15.Ap, 61.14.-x, 73.20.At

# I. INTRODUCTION

TiTe$_2$ is a prototype material in a large family of layered transition metal dichalcogenides (TMDCs) whose quasi-2-dimensional (quasi-2D) structural and electronic properties have been intensively studied during the past few decades (for a recent review see, for example, Ref. 1). TiTe$_2$ crystallizes in the 1T-CdI$_2$ structure, which is characterized by tightly bound chalcogen-metal-chalcogen layers separated by a Van-der-Waals gap. Strong intra-layer and weak inter-layer bonding in such a structure result in highly anisotropic quasi-2D properties of TiTe$_2$ characterized, for example, by a ratio of the out-of-plane to in-plane resistivity as much as 35-40, a value typical of the TMDCs. The electronic structure of TiTe$_2$ is formed by partial overlap of the Te 5$p$ derived valence states with the Ti 3$d$ derived conduction states, which results in a semimetallic behavior of this material. Due to small electron-phonon coupling parameter ($\lambda \sim 0.22$) it does not seem to exhibit superconductivity or charge-density-wave instabilities typical of other quasi-2D materials.

Extensive angle-resolved photoemission (PE) experiments on TiTe$_2$ have delivered a good knowledge of its band structure $E(\mathbf{k})$ with resolution in the $\mathbf{k}$-space (for a few recent entries see Refs. 2-5). The PE electron removal spectra are relevant for the material properties because they reflect the hole spectral function $A(\omega,\mathbf{k})$ weighted with the PE matrix element. Of particular interest are such studies on the Fermi surface (FS) immediately connected to the transport properties. For example, the Ti $3d_{z^2}$ band forming an electron pocket of the FS around the ML line of the Brillouin zone (BZ) has been exploited as a playground to test the Fermi liquid theory.[2-5] However, such studies analyzed the PE data mostly with respect to $E(\mathbf{k})$ as a function of the layer-parallel (with the natural cleavage plane of TMDCs, surface-parallel) wavevector component $\mathbf{k}_{//}$, remaining thus basically within a 2D view of the electronic structure.

3-dimensional (3D) effects in the electronic structure of TiTe$_2$ arise from the inter-layer interactions. They are expressed by $E(\mathbf{k})$ as a function of the layer-perpendicular (surface-perpendicular) wavevector component $k_\perp$. Despite the quasi-2D nature of TiTe$_2$, the 3D effects are significant. Firstly, PE spectra measured with variable photon energy $h\nu$ suggest that the $E(k_\perp)$ dispersion can reach a range of ~2.5 eV over the BZ extension.[2,6] The available band calculations yield similar figures. Furthermore, the very fact of non-zero out-of-plane conductivity suggests existence of bands having non-zero layer-perpendicular group velocity $\partial E/\partial k_\perp$ at the FS. Recent high-resolution PE studies[3,5] have found that even the model Ti $3d_{z^2}$ derived FS pocket shows a residual PE linewidth of 14-17 meV surviving in the limit of negligible electron-phonon and electron-electron scattering. Extrapolated to the limit of negligible impurity scattering,[3] this figure suggests a residual 3D dispersion with $\partial E/\partial k_\perp$ of the order of 0.12 eV·Å.

$\mathbf{k}$-resolved studies of the 3D effects in TiTe$_2$ suffer from a fundamental difficulty of the PE experiment: In a simplified picture of the PE process, which includes photoelectron excitation within the crystal bulk and its escape into vacuum, $k_\perp$ is conserved at the excitation stage but gets distorted at the escape stage. The information on the initial-state $k_\perp$ can however be recovered if the final-state surface-perpendicular dispersion $E(k_\perp)$ back into the crystal bulk is known. A common solution to this problem is the use of empirically adjusted free-electron-like (FE-like) final states. However, for TiTe$_2$ such an approach fails. This is clear, for example, from the results of PE band mapping of the valence-band $E(k_\perp)$ from FE-like final states:[2] the resulting experimental points are highly inconsistent. Another example is broadening of PE peaks at low energies:[4] under assumption of FE-like final states it shows an energy dependence opposite to the trend predicted by the mean free path 'universal curve'. These facts evidence that the final states in TiTe$_2$, similarly to other quasi-2D materials,[7-9] feature dramatic *non-free-electron* (non-FE) *effects* – deviations from the FE-like approximation resulting from photoelectron multiple scattering by the crystal potential. Furthermore, the final states can experience significant energy shifts due to band- and $\mathbf{k}$-dependent excited-state self-energy corrections $\Delta\Sigma$.[10]

Another aspect of the PE experiment is that the finite photoelectron lifetime damps the final-state wavefunction in the surface-perpendicular direction towards the crystal interior.[11,12] Such a confinement results in intrinsic final-state broadening in $k_\perp$. The PE peaks then reflect not exactly the initial-state $E(k_\perp)$ dispersion, but its average over the broadening interval.

Interpretation of the PE data from TiTe$_2$ with respect to the valence band $E(k_\perp)$ requires therefore knowledge of the true final-state $E(k_\perp)$ dispersions, including the non-FE and self-energy effects, and the final-state lifetimes describing its damping. This can be achieved with an independent Very-Low-Energy Electron Diffraction (VLEED) experiment. The relevance of VLEED to PE is based on the one-step PE theory which, neglecting the electron-hole interaction, treats the final states as the time-reversed LEED states (see, for example, Ref.11). In the VLEED spectra of elastic reflectivity $R(E)$, energies of the spectral structures reflect the characteristic points in the final-state $E(k_\perp)$ such as the band gap edges, and their broadening and relative amplitudes reflect the corresponding lifetimes (see Refs. 13,14 and references therein).

Here, we present a study of 3D effects in TiTe$_2$ using a combination of the VLEED and angle-resolved PE spectroscopies. The study focuses on the ΓA direction of the BZ. First, the final-state $E(k_\perp)$ dispersions and lifetimes are determined from the VLEED experiment supported by first-principles calculations of complex band structure. The non-FE effects in the final states such as non-parabolic dispersions and multiband composition are analyzed in detail. Second, the valence band $E(k_\perp)$ is determined to a great detail from extensive $h\nu$-dependent PE experimental data interpreted using the VLEED derived final states. The full control over the 3D wavevector achieved with such a combination of experimental techniques yields new findings about the electronic structure of TiTe$_2$ such as the absence of the Te $4p_z$* electron pocket of the FS.

## II. UNOCCUPIED STATES

### A. VLEED experiment and results

Our experimental technique is described in detail elsewhere.[15,16] Briefly, we used a standard four-grid LEED optics operating in the retarding field mode: The electrons are first accelerated in the gun to energies around 300eV to form a well-focused beam, and then decelerated to the required primary energy $E$ in a retarding field between the gun and the sample, maintaining focusing down to the lowest energies. For the angle-dependent measurements, distortion of the off-normal electron trajectories and thus incident surface-parallel wavevector $\mathbf{K}_{//}$ due to the retarding field was taken into account as described in Ref. 15. The VLEED spectra were measured as the elastic electron transmission spectra $T(E)$, which are related to the total elastic reflectivity $R(E)$ integrated over all diffracted beams as $T(E)=1-R(E)$. The measurements were performed in the target current circuit, taking advantage of fairly structureless inelastic reflectivity contribution to the target current. The energy spread of the primary electrons was ~0.25eV HWHM. Atomically clean surface of TiTe$_2$ was obtained by usual in-situ cleavage. The workfunction was determined to be 5.5±0.2 eV.

The experimental angle-dependent $T(E)$ spectra are shown in Fig. 1. They were measured under $\mathbf{K}_{//}$ variation along the $\overline{\Gamma\mathrm{K}}$ azimuth of the surface BZ (ΓAHK plane of the bulk BZ). Corresponding energy dependences of the surface-parallel incident wavevector $\mathbf{K}_{//}$ are shown in the inset. With each spectrum taken at a fixed sample rotation angle, the retarding field increases the incident angle towards lower energies, reducing energy variations of $\mathbf{K}_{//}$ along the spectrum compared to the field-free case.[15]

The experimental $T(E)$ spectra show prominent structures dispersing with $\mathbf{K}_{//}$. In a series of previous works (see, for example, Refs. 7,14,16 and the references therein) it has been established that the VLEED spectral structures reflect unoccupied 3-dimensional $E(\mathbf{k})$ along the BZ direction(s) defined by $\mathbf{K}_{//}$ conservation. Of all bands available for given incident $E$ and $\mathbf{K}_{//}$, only so-called *coupling* (or *conducting*) *bands* – whose wavefunctions allow effective coupling to the incident plane wave and electron transport into the crystal – are involved in formation of the VLEED spectrum. Specifically, the $T(E)$ sharp changes (identified as the $dT/dE$ extremes) reveal in the $E(k_\perp)$ dispersions of these bands the critical points such as the band edges. In view of further implications of VLEED for analysis of the PE data, it is important to note that the same coupling bands effective in VLEED are effective as the final bands in the PE process[14,17] (see II.C.1). Further, the coupling bands in the PE context will be synonymously referred to as the final bands.

$\mathbf{K}_{//}$ dispersion of the experimental spectra is represented in Fig. 2. Here, the energy intervals within the $T(E)$ minima (identified by $d^2T/dE^2 > 0$) in each spectrum are represented by white areas in ($K_{//}$, $E$) coordinates, and the intervals within the $T(E)$ maxima ($d^2T/dE^2 < 0$) by shaded areas. With the above physical meaning of the $T(E)$ structures, the borders between these areas directly represent $E(\mathbf{k}_{//})$ surface projected dispersion of the critical points in final-state $E(k_\perp)$ dispersion for given $\mathbf{K}_{//}$.

Apart from the gross $T(E)$ structures due to the bulk $E(\mathbf{k})$, weak narrow oscillations can be distinguished in the $\mathbf{K}_{//}$ dispersion map. They are placed near the diffraction thresholds $E=\hbar^2(\mathbf{K}_{//}+\mathbf{g})^2/2m$ marked in Fig.2 by dashed lines corresponding to different $\mathbf{g}$. Such oscillations manifest the surface resonance (SR) states formed by the pre-emergent diffraction beam traveling along the surface through multiple reflections between the surface barrier and the crystal bulk (physics of the SR states is discussed in detail in Refs.14,18,19). The multiple scattering mechanism makes the SR states similar in origin to the image potential states, but the reflection phases on the crystal side are different because of different energies falling above the vacuum level and $\mathbf{K}_{//}$ associated with the surface-parallel movement of the beam. With the wavefunctions concentrated outside the bulk crystal, the SR states experience weaker potential corrugations and show almost free-electron dispersion. Their presence, highly sensitive to the surface contamination, indicates excellent surface quality. SR phenomena have also been observed for other quasi-2D materials such as $TiS_2$.[20]

## B. Computational procedure and results

### 1. Complex band structure and VLEED spectra

Reference calculations of the VLEED spectra and corresponding unoccupied $E(\mathbf{k})$ are crucial for interpretation of the experimental VLEED data in terms of band structure. We focus on the normal-incidence data reflecting $E(\mathbf{k})$ along the ΓA direction of the BZ. The cornerstone of our computational approach is the Bloch wave formalism of the multiple scattering LEED theory (see, for example, the seminal works in Refs. 21-23). In this formalism, the electron scattering in the VLEED experiment is described by matching the electron wavefunction in the vacuum half-space $\Phi_{vac}(\mathbf{r})$ (= a superposition of the plane waves corresponding to the incident and all diffracted beams) to that in the crystal half-space $\Phi_c(\mathbf{r})$ (= a superposition $\sum_{\mathbf{k}} A_{\mathbf{k}} \phi_{\mathbf{k}}(\mathbf{r})$ of the Bloch waves $\phi_{\mathbf{k}}(\mathbf{r})$ excited in the crystal) under conservation of $E$ and $\mathbf{k}_{//}$. The set of $\phi_{\mathbf{k}}(\mathbf{r})$ in the semi-infinite crystal is determined from the Schrödinger equation

$$\left[-\frac{\hbar^2}{2m}\Delta + V(\mathbf{r}) - iV_i - E\right]\phi_{\mathbf{k}}(\mathbf{r}) = 0,$$

where $V(\mathbf{r})$ is the crystal potential and the inelastic scattering is included through the spatially constant absorption potential $V_i$ connected to the electron lifetime as $V_i=\hbar/\tau$. In the elastic limit $V_i=0$, the $\phi_{\mathbf{k}}(\mathbf{r})$ solutions are either propagating into the crystal interior (*bulk* Bloch waves, having real $k_\perp$) or damped in this direction (*surface* ones, having complex $k_\perp$ with $\mathrm{Im}\,k_\perp$ reflecting the damping rate). With $V_i \neq 0$, all $\phi_\mathbf{k}(\mathbf{r})$ become damped into the crystal interior, and are described on equal footing by complex $k_\perp$ (note however that by virtue of the surface-parallel invariance of the LEED process $\phi_\mathbf{k}(\mathbf{r})$ are undamped along the surface and have real $\mathbf{k}_{//}$). The corresponding $E(\mathbf{k})$ is the *complex band structure* in the sense of real $E$ depending on complex $k_\perp$. Calculation of the complex $E(\mathbf{k})$ and corresponding $\phi_\mathbf{k}(\mathbf{r})$ is an *inverse* band structure problem: Given $E$ and $\mathbf{k}_{//}$, the secular equation is solved for complex $k_\perp$ values. Computationally, this is the most demanding part of the Bloch waves formalism.

A novel feature of the present computational scheme is incorporation of a realistic crystal potential in the surface region, where it strongly deviates from the periodic potential in the bulk. The LEED states are calculated with the recently developed embedding method,[24] within which two self-consistent calculations are performed: (i) for the infinite $TiTe_2$ crystal, and (ii) for the surface region, which is a fragment of a periodic slab with the unit cell containing three Te-Ti-Te three-layers (nine atomic layers) and a vacuum region to separate the three-layers from each other. The former yields the self-consistent $V(\mathbf{r})$ in the bulk, and the latter that in the surface region matching the bulk potential already in the outermost layer. Fig. 3 shows the obtained potential distribution in the surface region. The slab is thick enough so that the interaction between the slabs is negligible: the $V(\mathbf{r})$ profile in the middle of the slab coincides with that in the bulk crystal, and in the vacuum region it grows to reach a constant value of 5.6 eV, which is in good agreement with our VLEED experimental workfunction. To calculate the LEED state, a fragment is cut out of the slab unit cell and embedded between the bulk crystal half-space on the left side and the vacuum half-space with a constant potential on the right side. In the embedded region and vacuum half-space $V_i$ is set to zero. In the bulk half-space the LEED function is a superposition of the inverse band structure solutions $\phi_\mathbf{k}(\mathbf{r})$, and in the embedded region the function is expanded in terms of the eigenfunctions of the repeated slab Hamiltonian for a given $\mathbf{k}_{\|}$. In the vacuum region the LEED wave function is represented by the incoming plane wave and reflected plane waves (including decaying ones) corresponding to all surface reciprocal vectors $\mathbf{g}$. The solution, a smooth function satisfying the Schrödinger equation in the embedded region and, by construction, in the crystal and vacuum half-spaces, is obtained by a variational method described in detail in Ref. 24.

Further details of our computational methodology have been presented in Refs. 16,24,25. Briefly, the standard Density Functional Theory (DFT) formalism with the Local Density Approximation (LDA) exchange-correlation is used. Both the self-consistent and scattering calculations are performed with the Extended Linearized Augmented Plane Waves (ELAPW) method. The radial basis sets for the lower angular momenta were extended by additional functions to ensure high accuracy of the wavefunctions over the energy region up to 50 eV above $E_F$. The inverse band structure problem for complex $k_\perp$ is solved using an exact $\mathbf{k}\cdot\mathbf{p}$ method using a basis set of bulk band structure wavefunctions. This method allows computationally efficient reduction of the inverse band structure problem to a linear algebra eigenvalue problem. The energy dependence of $V_i$ is determined empirically by fitting the energy broadening and relative amplitudes of the experimental spectral structures.[16,26]

The theoretical normal-incidence $T(E)$ spectrum in comparison with the experimental one is shown in Fig. 4. The excellent agreement with the experiment regarding the positions, energy broadenings and relative amplitudes of all spectral structures proves the relevance of our theoretical framework of the VLEED process in application to $TiTe_2$. The correct description of the potential distribution in the surface region has turned out to be crucial to achieve accurate theoretical $T(E)$: The approximation of a step-like surface barrier, which was found sufficient for $NbSe_2$ and graphite,[16,26] introduced a considerable error for $TiTe_2$.

The inset in Fig. 4 shows the $V_i(E)$ energy dependence used in the calculations. It was estimated by varying $V_i$ to fit the energy broadenings and relative amplitudes of the spectral structures in the experimental normal-incidence $T(E)$ spectrum. With the actual sensitivity of the theoretical spectra to the variations of $V_i$, it was sufficient to judge the quality of the fit by eye in order to estimate $V_i$ to within ±20 %. The $V_i(E)$ dependence shows a sharp increase in the low-energy region. Our supplementary ab initio calculation of the dielectric function in the random phase approximation yield the bulk plasmon energy $\hbar\omega_p$ around 18 eV, in agreement with the available experimental data.[27] This suggests that the main contribution to the $V_i(E)$ increase is due to excitation of the bulk plasmon. The characteristic plasmon step[26] is however not resolved in our $V_i(E)$ dependence, within the accuracy of our $V_i$ evaluation.

The theoretical unoccupied $E(\mathbf{k})$ along $\Gamma A$, underlying the normal-incidence $T(E)$ calculations, is shown in Fig. 5 (*a,b*). Reflecting the damped nature of the $\phi_\mathbf{k}(\mathbf{r})$ Bloch waves involved in the VLEED process, this is the complex band structure in the sense of real $E$ depending on complex $k_\perp = \text{Re}k_\perp + i\text{Im}k_\perp$. Of all $\phi_\mathbf{k}(\mathbf{r})$ generated by our calculations, the figure shows only those characterized by a small damping rate $\text{Im}k_\perp < |\Gamma A|$. Note that with $V_i \neq 0$ the complex band structure is radically different from that in the $V_i = 0$ limit: The $E(\text{Re}k_\perp)$ dispersions pass through the band gaps continuously, with the critical points surviving only as distinct changes in the dispersion slope.[13] The band gaps are better distinguished in the $E(\text{Im}k_\perp)$ plot as loop-like enhancements of $\text{Im}k_\perp$.

## *2. Partial absorbed currents*

Identification of the coupling bands, dominating in the VLEED and PE processes, employed a calculation of *partial absorbed currents* $T_\mathbf{k}$ characterizing the partial contributions of each $\phi_\mathbf{k}(\mathbf{r})$, constituting the total LEED state in the crystal $\Phi_c(\mathbf{r}) = \sum_\mathbf{k} A_\mathbf{k} \phi_\mathbf{k}(\mathbf{r})$, to the $T(E)$ total absorbed current. In the $V_i = 0$ elastic case, $T_\mathbf{k}$ are calculated with the usual expression $T_\mathbf{k} = -i|A_\mathbf{k}|^2 \frac{\hbar}{2m}(\phi_\mathbf{k}^* \frac{\partial}{\partial r_\perp}\phi_\mathbf{k} - \phi_\mathbf{k}\frac{\partial}{\partial r_\perp}\phi_\mathbf{k}^*)$ for the propagating $\phi_\mathbf{k}(\mathbf{r})$, and $T_\mathbf{k} = 0$ for the damped ones.

In the $V_i \neq 0$ case all $\phi_\mathbf{k}(\mathbf{r})$ become damped, and the usual concept of the currents associated with propagating wavefunctions collapses. In Refs. 13,17 it was shown that the current absorbed in the crystal appears in this case due to the electrons inelastically scattered away from the coherent wavefunction, and can be calculated by integrating the density of the LEED state over the crystal half-space as

$$T = \frac{2V_i}{\hbar}\int_\Omega |\Phi_c(\mathbf{r})|^2 d\mathbf{r}$$

(with this generalization, the current conservation theorem was utilized in our calculations as a criterion of the computational accuracy: the sum of the currents carried by the incident and diffracted beams, i.e. the current in the vacuum halfspace, must be equal to the current $T(E)$ absorbed in the crystal halfspace). The $T_\mathbf{k}$ partial absorbed currents are then defined as

$$T_\mathbf{k} = \frac{2V_i}{\hbar}\int_\Omega |A_\mathbf{k}\phi_\mathbf{k}(\mathbf{r})|^2 d\mathbf{r}$$

In the $V_i \rightarrow 0$ limit this expression reduces to the usual elastic current. Physically, large $T_\mathbf{k}$ values are characteristic of $\phi_\mathbf{k}(\mathbf{r})$ which (1) effectively couple to the incident plane wave and thus receive large excitation amplitudes, and (2) penetrate deep into the crystal to enable effective electron transport. In the free-electron case one Bloch wave, identical to the incoming plane wave, receives $T_\mathbf{k} = 1$ and all others $T_\mathbf{k} = 0$.

In contrast to the $V_i = 0$ case, with $V_i \neq 0$ surface-related $\phi_\mathbf{k}(\mathbf{r})$ can acquire non-zero $T_\mathbf{k}$ and contribute to the total current similarly to the bulk-related ones. Owing to the interference terms $\int_\Omega A_{\mathbf{k'}}A_\mathbf{k}\phi_{\mathbf{k'}}(\mathbf{r})\phi_\mathbf{k}(\mathbf{r})d\mathbf{r}$

between the $\phi_\mathbf{k}(\mathbf{r})$ constituents of total $\Phi_c(\mathbf{r})$ in the expression for $T(E)$, the sum of $T_\mathbf{k}$ is not exactly equal to total $T(E)^{17,26}$ and $T_\mathbf{k}$ values can even exceed 1.

In Fig. 5 (*a*) the calculated $T_\mathbf{k}$ values for each band are shown in grayscale. The bands with strong damping $|\mathrm{Im}k_\perp|>|\Gamma A|$ are omitted from the plot, because they anyway have vanishing $T_\mathbf{k}$.

### C. Properties of the final states

#### *1. Role of the $T_k$ partial absorbed currents in VLEED and PE*

The unoccupied $E(\mathbf{k})$, due to the progressive increase with energy of the number of bands folding into the reduced BZ, always appears as a multitude of bands. This general fact is illustrated by our results for TiTe$_2$ in Fig. 5 (*a,b*). Among all bands, however, only a few coupling bands identified by sufficient $T_\mathbf{k}$ values (these bands are labeled *1-3*) make significant contributions to the $T(E)$ total absorbed current and thus form the $T(E)$ spectrum. The majority of other bands in the multitude available for given $E$ and $\mathbf{k}_{//}$ are characterized by vanishing $T_\mathbf{k}$ values and are thus ineffective in the VLEED process.

By virtue of the time-reversal relation between the VLEED state and PE final state, the latter is composed of the same $\phi_\mathbf{k}(\mathbf{r})$. In analogy with the VLEED process, the partial contribution of each $\phi_\mathbf{k}(\mathbf{r})$ to the total photocurrent $I(E)$ can be characterized by *partial photocurrents* $I_\mathbf{k}$ defined in the framework of the one-step PE theory as

$$I_\mathbf{k} \propto \left| < A_\mathbf{k}^* \phi_\mathbf{k}^*(\mathbf{r}) | \mathbf{A} \cdot \mathbf{p} | \Psi(\mathbf{r}) > \right|^2,$$

where $\mathbf{A}$ is the electromagnetic field vector potential, $\mathbf{p}$ the momentum operator, and $\Psi(\mathbf{r})$ the initial-state wavefunction. In Ref.17 it was shown that $I_\mathbf{k}$ are in fact proportional to $T_\mathbf{k}$ such that

$$I_\mathbf{k} \propto |M_{fi}(\mathbf{k})|^2 \, T_\mathbf{k},$$

where $M_{fi}$ is the phototransition matrix element (involving the oscillating part of the final-state $\phi_\mathbf{k}(\mathbf{r})$ to characterize the phototransition process decoupled from the photoelectron escape[11]). Therefore, the coupling bands dominating in VLEED also dominate as the final bands in PE (if not impeded by vanishing $M_{fi}$). Physically, the states effective in coupling to the incoming plane wave and electron transport into the crystal in VLEED are equally effective in photoelectron transport out of the crystal and coupling to the outgoing plane wave in PE. The coupling bands in our unoccupied $E(\mathbf{k})$ in Fig. 5 (*a*) are thus exactly the final bands in the PE process.

#### *2. Non-free-electron effects*

The final-state bands in TiTe$_2$, represented by the $E(Rek_\perp)$ panel of Fig. 5 (*b*), demonstrate dramatic non-FE effects:

(i) Within the FE-like approximation, implying spatially constant crystal potential, the PE final state with energy $E_f$ includes only one coupling band (whose surface-parallel wavevector matches $\mathbf{K}_{//}$ in vacuum) whereas $T_\mathbf{k}$ of all other bands are strictly zero.[17,28] Fig. 5 (*a*) shows that for TiTe$_2$ this is not the case: through the whole energy range the final state comprises multiple (mostly two) coupling bands having comparable $T_\mathbf{k}$ – and thus $I_\mathbf{k}$ partial photocurrent – magnitudes. As we will illustrate in III.C.1, multiple $Rek_\perp$ available in such a *multiband* final state result in multiple PE peaks corresponding to direct transitions at different $Rek_\perp$.

Multiband composition of the final state is in fact a typical non-FE effect, in conventional terms of PE spectroscopy often referred to as umklapp (see, for example, Ref. 29) or secondary cone emission.[30,31] Note that this effect must include band hybridization in the final state.

The multiband composition significantly complicates the relation of the final states to the VLEED spectra. In the case of one coupling band this relation is simple: the band gap regions manifest themselves as the $T(E)$ minima and the regions of smooth dispersion between them as the $T(E)$ maxima, with the critical points corresponding to the $dT/dE$ extremes (see, for example, Refs. 8,14). To illustrate this relation in our case, in Fig. 5 (*c*) we reproduce the calculated $T(E)$ spectrum together with $dT/dE$. One interesting region is around 15 eV, where one of the two bands forming the spectrum, the band *1*, passes through a band gap. Its $T_\mathbf{k}$ decreases here due to enhanced damping of the corresponding $\phi_\mathbf{k}(\mathbf{r})$. Surprisingly, the total $T(E)$ shows here a maximum. This occurs due to even stronger increase of $T_\mathbf{k}$ in another coupling band, the band *2*. Physically, this increase can be explained by that the enhanced damping of the band *1* reduces the strengths of its hybridization with the band *2* determined by overlap of the corresponding $\phi_\mathbf{k}(\mathbf{r})$; as a result, the latter gets closer to the free-electron character and yields larger $T_\mathbf{k}$. This example illustrates that in the multiband case the band gaps are not necessarily manifested by $T(E)$ minima. Another interesting region is around 31 eV. Despite $T_\mathbf{k}$ of both bands *1* and *2* reach here a maximum, the total $T(E)$, surprisingly, shows a minimum. This fact is attributed to interference between the two $\phi_\mathbf{k}(\mathbf{r})$ so that the sum of $T_\mathbf{k}$ is not exactly equal to the total $T(E)$ (see II.B.2). To embrace such cases, the relation of the multiband states to VLEED can be best understood with reference to the $dT/dE$ spectra: the critical points in the coupling bands, where the Bloch waves composition undergo sharp changes, always correspond to the extremes (minima or maxima) in $dT/dE$;

(ii) While the FE-like dispersion is parabolic, in TiTe$_2$ each of the final bands strongly deviates from such a behavior in the regions where it experiences hybridization with other bands (which would form the band gaps in the $V_i$=0 band structure). The FE-like approximation with the inner potential $V_{000}$ and effective mass $m_0$ as adjustable parameters will describe such dispersions only as a local fit with $V_{000}$ and $m_0$ strongly depending on $E$ and $\mathbf{k}$.

The non-FE effects are also apparent immediately in the VLEED experimental surface projection of the final states in Fig. 2. The $E(\mathbf{k}_{//})$ dispersions remain parabolic only over limited $E$ and $\mathbf{k}_{//}$ intervals, with clear discontinuities between them indicating discontinuities in the $V_{000}$ and $m_0$ parameters of the FE-like fit. The FE-like approximation thus fails to describe the final bands of TiTe$_2$.

Non-FE effects in the final states, including their multiband composition, are in fact prominent for the quasi-2D materials due to strong modulation of the crystal potential in the layer-perpendicular direction.[8,9,32] However, in certain $E$ and $\mathbf{k}_{//}$ ranges such effects were observed even for Cu[14,17] and Al,[33] the materials conventionally considered to have purely FE-like final states, Bi,[31] GaAs,[34] GaN,[35] etc.

### 3. Final-state $k_\perp$ broadening

Any final-state $\phi_\mathbf{k}(\mathbf{r})$, a damped Bloch wave with complex $k_\perp$, can be represented as a superposition of propagating waves with real $k_\perp$. These $k_\perp$ form a Lorentzian distribution centered at $Rek_\perp$ and having a full width of $2Imk_\perp$.[12,36] Thus, $\delta k_\perp = 2Imk_\perp$ represents the *final-state broadening in $k_\perp$*. This aspect of the final bands in TiTe$_2$ is reflected in the $E(Imk_\perp)$ panel of Fig. 5 (*a*). The following properties of $\delta k_\perp$ are observed:

(i) $\delta k_\perp$ is different for each band and undergoes significant energy variations. In particular, it shows loop-like enhancements corresponding to the band gaps in the $V_i$=0 band structure, where the Bloch waves experience additional damping due to scattering off the crystal potential (see, for example, Refs. 13,23,26);

(ii) Outside the band gaps where $\delta k_\perp$ is predominantly due to inelastic scattering and connected with $V_i$ as $\delta k_\perp = V_i \frac{\partial \mathrm{Re} k_\perp}{\partial E}$,[23] the energy dependence of $\delta k_\perp$ follows that of $V_i$ (see inset in Fig. 4).

In conventional PE data analysis, $\delta k_\perp$ is estimated from the final-state energy broadening $\delta E_f$ of the PE peaks measured in the Constant-Initial-State mode with the initial state fixed at the Fermi level to reduce the initial-state lifetime contribution; the experimental $\delta E_f$ is then translated into $\delta k_\perp = \delta E_f \frac{\partial k_\perp}{\partial E}$.
However, such analysis does not take into account the non-FE effects in the final states. In particular, for TiTe$_2$ (Ref.4) it has yielded in a low $E_f$ region around 13 eV a $\delta k_\perp$ value of ~0.27 Å$^{-1}$, which is much too large compared to the 'universal curve' of the mean free path $\lambda$ mirroring $\delta k_\perp$ as $\lambda=1/\delta k_\perp$. Furthermore, this $\delta k_\perp$ value was found to decrease to ~0.15 Å$^{-1}$ when going to higher $E_f$ around 21 eV, whereas the 'universal curve' predicts an increase of $\delta k_\perp$ in this energy region. The origin of these inconsistencies is the multiband composition of the final states in TiTe$_2$ discussed above. Different final bands, see Fig. 5 (*a*), unresolved in this PE experiment in separate peaks merged in one peak with enhanced $E_f$ broadening, whose unwary translation into $\delta k_\perp$ resulted in overestimated $\delta k_\perp$ values. With increase of $E_f$ towards 21 eV energy separation of the two final bands somewhat decreases, explaining the apparent decrease of overall $\delta k_\perp$. An additional drawback of such an incautious analysis of the PE data is neglecting the enhancements of $\delta k_\perp = 2\mathrm{Im} k_\perp$ in the band gap regions discussed above.

### *4. Experimental energy corrections*

One-to-one correspondence of the spectral structures in the theoretical and experimental normal-incidence $T(E)$ in Fig. 4 allows a correction of the theoretical final states according to the VLEED experiment.

The values of the energy shifts $\Delta E$ between the experimental and theoretical d$T$/d$E$ extremal points are shown in Fig. 6. Assuming that the remnant computational inaccuracies in the matching procedure and underlying $E(\mathbf{k})$ are negligible, these shifts are fundamentally due to the excited-state self-energy corrections $\Delta\Sigma$, which appear due to the difference of the excited-state exchange-correlation potential from the ground-state one implied by our DFT based calculations (assuming that the LDA approximation is accurate). The observed $\Delta E$ energy dependence is non-monotonous, with a clear discontinuity near 10 eV where the dominant band forming the VLEED structures hops from *1* to *2*, an oscillation near 20 eV, and prominent increase starting from 30 eV. Such peculiarities are expected in view of the non-monotonous band and energy dependence of $\Delta\Sigma$.[10]

The experimental $\Delta E$ values, absorbing the self-energy effects and partly computational inaccuracies, were used to correct the energy position of the theoretical final states from Fig. 4 (*a*) according to the VLEED experiment. The energy correction was taken as a smooth curve obtained by Gaussian smoothing of the scattered $\Delta E$ values to remove kinks in the corrected $E(\mathrm{Re}k_\perp)$ dispersions. In the low-energy region the curves for the bands *1* and *2* were taken different, and above 15 eV the same curve characteristic of the band *2* was applied to all bands *1-3*. Our further PE analysis utilizes the corrected final states.

## III. VALENCE BANDS

### A. PE experiment

#### 1. Experimental procedure and results

The PE experiment was performed at the SA73 bending magnet beamline of the SuperACO storage ring at LURE, France. Synchrotron radiation polarized in the horizontal plane was incident at angle of 45° relative to the surface normal in the MΓM' azimuth. The spectra were measured at normal emission in the EDC mode with photon energies $h\nu$ varying from 11.5 to 33 eV in steps of 0.5 eV. The combined monochromator and analyser energy resolution varied from 23 meV at the lowest $h\nu$ limit to 130 meV at the highest one. The monochromator energy scale was calibrated with respect to the Fermi edge through all measured PE spectra assuming an analyser workfunction of 4.3eV. The spectra were acquired with statistics between ~$3\times10^4$ and $2\times10^5$ counts per energy window of ~33meV. Control spectra, taken after completing the measurement series ~50h after the cleavage, showed only insignificant background increase, indicating negligible surface contamination.

The raw EDC spectra $I(E)$ are shown in Fig. 7 (*left*). The achieved statistics allows clear resolution of finer spectral details such as merging peaks. The spectra (minus the intensity above $E_F$ due to high-order light) are normalized to the same integral intensity.

In Fig. 8 (*a, left*) the above normalized EDC spectra are rendered into a PE intensity map as a function of the final- and initial-state energies $E_f$ and $E_i=E_f-h\nu$ (relative to $E_F$). The individual EDCs taken at constant $h\nu$ are seen as inclined lines.

Fig. 8 (*b, left*) shows a similar map obtained from the negative second derivative $-d^2I/dE^2$ of the EDCs, with the negative values of $-d^2I/dE^2$ set to zero. Such a representation enhances the spectral structures including the peaks and shoulders[14] (except for the structures whose dispersion in the ($E_f,E_i$) coordinates is along the derivation direction, i.e. along the EDC lines). Due to noise enhancement in the second derivative, the original EDCs were denoised here by Gaussian smoothing using a halfwidth linearly increasing from 60 meV at the high-$E_i$ end to 240 meV at the low-$E_i$ end in accordance with increase of the spectral structures energy width. Due to large intensity variations, a logarithmic intensity scale is used in our $-d^2I/dE^2$ map.

#### 2. Overall picture of the PE spectral structures

The PE experiment was supported by calculations of the valence band $E(\mathbf{k})$ representing the bulk initial states of the PE process. These states, in contrast to the damped final states, are described by propagating Bloch waves and thus by real $k_\perp$. The calculations were performed within the standard DFT-LDA formalism using the self-consistent ELAPW method (see II.B.1). The scalar relativistic effects with the spin-orbit coupling in the second variation were included. The theoretical valence band $E(\mathbf{k})$ is shown in Fig. 9. Our results appear in general agreement with the previous calculations,[2,4] although there are some quantitative disagreements (see III.E).

Our spectra measured at normal emission reflect the $E(k_\perp)$ layer-perpendicular valence band dispersions along the ΓA direction. Comparison with the calculated valence band identifies the following origin of the principal spectral structures:

(i) The dominant peaks dispersing through the lower and upper part of the valence band originate from the bonding Te $5p_z$ and antibonding Te $5p_z^*$ states, respectively (note that in the $E_f$ region near 18 eV the Te

$5p_z^*$ peaks disappear in the -d$^2$I/dE$^2$ plot because they disperse along the derivation direction). As discussed in III.C.1, *multiple* dispersion branches of the Te $5p_z$ and Te $5p_z^*$ peaks reflect multiband composition of the final state. The PE dispersions in $E_f$ reflects dispersion of the Te $5p_z$ and $5p_z^*$ states in $k_\perp$ resulting from the layer-perpendicular orientation of the electron orbitals allowing their effective overlap in this direction;

(ii) The weak non-dispersive peak at $E_i$~1.7 eV, vanishing in the middle of our $E_f$ interval, results from the bonding Te $5p_{xy}$ states. The absence of dispersion in $E_f$ reflects vanishing dispersion of these states in $k_\perp$ resulting from the layer-parallel orientation of the electron orbitals minimizing their overlap in the layer-perpendicular direction. Due to excellent statistics of our experiment we for the first time clearly resolve *spin-orbit splitting* of the Te $5p_{x,y}$ band, as seen in the raw EDCs or -d$^2$I/dE$^2$ plot;

(iii) The weak narrow peak just below $E_F$ does not have any direct counterpart in calculated $E(\mathbf{k})$ along ΓA. When going away from the normal emission, it dramatically scales up (minimization of its amplitude was used in our experiment to adjust the normal emission angle) and disperses down in $E_i$ (see the off-normal PE data in Ref. 2). With the absence of dispersion in $h\nu$, such dispersion in $\mathbf{K}_{//}$ compared with calculated $E(\mathbf{k})$ suggest the origin of this peak as due to the antibonding Te $5p_{x,y}^*$ band. In principle, according to all available band calculations including ours, exactly at the ΓA line the Te $5p_{x,y}^*$ band comes slightly above $E_F$. However, already with small deviation in $K_{//}$~0.1 Å$^{-1}$ it disperses below $E_F$. In this case the Te $5p_{x,y}^*$ signal can mix into the normal-emission spectra due to non-zero angular acceptance of the analyzer (±1° HWHM) as well as certain planarity errors over the sample surface (of the order of ±1° HWHM, as estimated from angular spread of reflected white light beam on the chamber wall). The resulting combined $\Delta K_{//}$ spread varies from ±0.03 at the low-energy end of our $E_f$ interval to 0.07 Å$^{-1}$ at the high-energy end. Given the very large magnitude of the Te $5p_{x,y}^*$ signal,[2] such $\Delta K_{//}$ should be sufficient to built up a sizeable contribution to the normal-emission spectrum;

(iv) The peak marked by the vertical dashed line runs through the EDCs at constant $E_f$ independent of $h\nu$. This fact identifies its origin as due to secondary electron emission (SEE) excited by high-order light.

### B. PE computations

The PE spectra were calculated within the framework of one-step theory using a new Bloch waves based method developed within the ELAPW formalism.[37] Compared to the KKR based methods, it provides the most direct link of the PE spectra to the initial and final state band structure.

The final state of the PE process is treated as complex conjugate of the LEED state described in II.B. The initial state in the crystal half-space $\Psi_c(\mathbf{r})$ is a standing wave represented by a linear combination of Bloch waves $\Psi_c(\mathbf{r}) = \sum_\mathbf{k} B_\mathbf{k} \psi_\mathbf{k}(\mathbf{r})$ of semi-infinite crystal, including propagating and decaying waves. In the bulk asymptote $\Psi_c(\mathbf{r})$ retains only propagating waves incident from the crystal interior on the surface, plus a number of reflected waves traveling in the opposite direction (in case of the Te $5p_z$ and $p_z^*$ bands there is only one reflected wave for each incident wave).

The initial states are calculated by the following procedure. The AΓA extension of the BZ is divided into $N$ equal $k_\perp$ intervals (in the present calculation $N$=80). In the middle of each interval the direct band structure problem (determination of the energy eigenvalues for given $k_\perp$) is solved. The wavefunctions were ascribed to spin-orbit split states according to the expansion coefficients coming from the second-variation calculations (this approximation is plausible everywhere except the points near the bottom of the Te $5p_z^*$ band where it strongly hybridizes with the Te $5p_{xy}$ band). Of the $N$ Bloch waves $\psi_\mathbf{k}(\mathbf{r})$ yielded by this procedure within each band, N/2 ones with positive $\frac{\partial E}{\partial k_\perp}$ group velocity are selected. They represent $\psi_k(\mathbf{r})$ incident on the surface from the crystal interior. Then for each of the corresponding N/2 energies within each band the inverse complex band structure problem is solved. This yields the partial $\psi_k(\mathbf{r})$ – including decaying ones with complex $k_\perp$ – whose linear combination forms the $\Psi_c(\mathbf{r}) = \sum_k B_k \psi_k(\mathbf{r})$ initial-state wavefunction in the crystal half-space. To determine the $B_k$ coefficients, the scattering problem is solved with the embedding method of Ref. 24: In the embedded region (see Fig. 3) the initial-state wavefunction is a linear combination of the slab eigenfunctions for $\mathbf{k}_{//}=0$ and $k_\perp=0$ which matches (with high but finite accuracy) $\Psi_c(\mathbf{r})$ in the crystal half-space and matches (exactly) the linear combination of decaying plane waves in the vacuum half-space. As the Schrödinger equation in the crystal half-space is satisfied by construction, the problem reduces to obtaining the solution of the Schrödinger equation in the embedded region and in the vacuum half-space. This is done by minimizing the energy deviation $\|(\hat{\mathbf{H}} - E)\Psi\|$ within these two regions and simultaneously minimizing the function and derivative mismatch at the matching plane between the bulk crystal and the embedded region.

To calculate the momentum matrix elements between the initial and final states, the wavefunctions are expanded in plane waves using the gouging technique described in Ref. 38: Each atom is surrounded by a small sphere (in our case having a radius of 0.5 a.u., with the muffin-tin spheres radii of 2.57 a.u.) within which the wavefunction is forced to damp to reduce its oscillations. The resulting pseudo-wavefunction has a rapidly convergent plane wave expansion, which facilitates the matrix element calculations. The contribution from the small spheres is practically neglected within this approximation. However, the introduced error reduces faster than the third power of the gouging radius, and convergence of the results with respect to this parameter indicates negligible magnitude of the remnant error.

Using these matrix elements, the EDCs were calculated over a grid of 500 points through $E_i$ range from $E_F$ to -6 eV. With the initial-state energy broadening due to hole lifetime combined with the final-state $k_\perp$ broadening, the PE intensity at given $E_i$ collects contributions from each $k_\perp$ point through the BZ. It is therefore calculated as integral over the $k_\perp$ extension of the BZ. The integration was performed by summation over all our $k_\perp$ intervals assuming linear $E(k_\perp)$ dispersion and constant wavefunctions within each interval. The initial-state broadening was represented by Lorentzian with a variable FWHM, which grew linearly from 0.05 eV at $E_F$ to 0.4 eV at the valence band bottom.

The results of our PE calculations are shown in Figs.7-8 (*right*) represented in parallel with the experimental data. Comparison with their experimental counterparts (*left*) demonstrates remarkable agreement, in particular on dispersions of the spectral structures. The systematic displacements in $E_f$ and $E_i$ are due to the $\Delta\Sigma$ renormalization of the final state and valence band dispersions (see III.E). Notable flattening of the Te $5p_z^*$ bottom compared to the experiment is caused by too much hybridization with the Te $5p_{xy}$ band, whose calculated energy location is in the experiment pushed by the $\Delta\Sigma$ corrections to higher $E_i$ away from the Te $5p_z^*$ bottom. To the best of our knowledge, the achieved level of agreement with the experiment is the best among all one-step PE calculations on layered materials reported so far. The key element of the calculations has been the use of accurate final states.

Certain disagreements between the calculated and experimental spectral structures in intensity and shape may trace back to the basic approximations of the one-step PE theory such as using Kohn-Sham solutions for quasi-particle wavefunctions, neglecting the interaction of the photoelectron with the hole left behind, and describing the inelastic effects with damped coherent waves. At the present theoretical level the validity of this basically one-particle approach cannot be judged a priori, and the comparison with the experiment is the only way to feel its limitations. The achieved excellent description of the PE peak dispersions allows us to assume that the quasiparticle band structure is treated by the present theory correctly (to within the self-energy shift). We therefore ascribe the failure to reproduce the exact intensities and shapes of the spectral structures to deficiencies of the one-particle description of the photoemission process.

Our further analysis of relation between the PE spectra and initial- and final-state band structure will employ a *direct transitions (DT) plot* which shows the ($E_f$,$E_i$) positions of the PE peaks dictated by momentum conservation between the initial and final states. Such plot constructed for TiTe$_2$ with the theoretical initial bands from Fig. 9 and final bands from Fig. 5 (*b*) (including only the coupling bands *1-3* whose significant $T_\mathbf{k}$ values enable effective coupling to the outgoing photoelectron plane wave) is shown in Fig. 8 (*b,c*) superimposed on the calculated PE intensity. In general the PE peaks follow the DT lines (note that in some ($E_f$,$E_i$) regions the predicted PE peaks can disappear due to vanishing $M_{fi}$ matrix element). There are however notable deviations which trace back essentially to the final-state $k_\perp$ broadening (see III.C.2). In this respect the PE calculations allow testing the limits of the DT model in application to 3D band dispersions.

### C. Final-state effects

#### *1. Signatures of non-FE final-state dispersions*

With FE-like final states, including one single band coupling to vacuum, any valence band dispersing in $k_\perp$ manifests itself as one single PE peak whose $E_i$ as a function of $E_f$ displays characteristic regular oscillations between the $E(k_\perp)$ extreme energies. This is illustrated in Fig. 10 (*a*) which shows the DT plot for TiTe$_2$ derived from the theoretical Te 5$p_z$ and 5$p_z^*$ valence bands in Fig. 9 as the initial states and FE-like final states. The dispersion ranges of the valence bands were adjusted to fit those of the experimental PE peaks in order to incorporate the ΔΣ renormalization in the valence band (see III.E). The final states employed free-electron $m_0$ and $V_{000}$=14.5 eV, as empirically determined in Ref. 2 (in close agreement with a value of 14.0 eV from Ref. 4).

Comparison with the experimental PE data shows that these FE-like final states have some relevance in a limited interval of $E_f$ between 17.5 and 27.5 eV, but severely fails through the rest of the experimental $E_f$ range. Interestingly, the optimal description of our normal-emission data with free-electron $m_0$ is achieved with $V_{000}$ of ~2.1 eV *above* the vacuum level, the dashed line in Fig. 10 (*a*); such an anomalous value demonstrates the purely empirical character of the FE-like approximation. Although such empirical adjustments of $V_{000}$ and $m_0$ can reduce the discrepancies, the FE-like final states inherently fail to explain the multiple dispersion branches of the Te 5$p_z$ and 5$p_z^*$ peaks. Furthermore, the experimental PE dispersions appear notably distorted compared to those expected from the FE-like approximation.

Incorporation of the non-FE and self-energy effects into the final states radically improves the description of the experimental PE data. Fig. 10 (*b*) shows the DT plot for TiTe$_2$ derived from the same initial states, but with the final states replaced by the VLEED derived ones, the final bands *1-3* from theoretical unoccupied $E(\mathbf{k})$ in Fig. 5 (*b*), with the experimental energy corrections from Fig. 6.

The DT plot constructed from the VLEED derived final states reproduces most of the peculiarities of the experimental data. In particular, the multiple dispersion branches of the Te 5$p_z$ and 5$p_z^*$ peaks find their natural explanation as originating from the multiple bands *1-3* with different $k_\perp$ composing the final state (energy position of the band *3* may be less accurate because in the VLEED spectrum its signal was obscured by the bands *1* and *2* having larger $T_\mathbf{k}$). The experimental peak dispersions are also well reproduced. Such a radical improvement over the FE-like final states conjecture proves that the non-FE behavior of the final states is crucial for correct interpretation the of PE spectra of TiTe$_2$. Furthermore, the incorporated VLEED experimental corrections to the final state energies (Fig. 6) are seen to considerably improve the agreement with the experiment in the low- and high-$E_f$ regions; in the low-$E_f$ one the PE data clearly confirms the $\Delta\Sigma$ difference between the bands *1* and *2*. These facts demonstrate that the self-energy renormalization of the final state dispersions is also important for interpretation of the PE data for TiTe$_2$.

Deviations of the experimental peaks from the DT plot near the top of the Te 5$p_z^*$ band (note that near its bottom such deviations disappear) indicate flattening of its $E(k_\perp)$ dispersion towards $E_F$ (see III.E). In the $E_f$ region near 17 eV within the Te 5$p_z$ band the peaks from the branches *1* and *2* merge into one, and in an extended $E_f$ region around 23 eV the peaks from all three branches merge. This results from large broadening of the peaks in $E_i$ due to decrease of hole lifetime when going towards deeper $E_i$. The remaining minor discrepancies between the experimental data and DT plot are mostly the intrinsic shifts due to the final-state $k_\perp$ broadening discussed in III.C.2.

Our PE calculations in Fig. 8 (*right*) incorporate the same non-FE final states (without the VLEED experimental corrections). They reproduce therefore all peculiarities of the PE data including the multiple dispersion branches.

### *2. Signatures of final-state $k_\perp$ broadening: Intrinsic shifts and 1DOS peaks*

On a qualitative level, the final-state $k_\perp$ broadening (see II.C.3) causes the PE signal to represent an average of the $M_{fi}$ weighted initial-state $E(k_\perp)$ dispersion over the $\delta k_\perp$ interval. In this respect $\delta k_\perp$ appears as intrinsic $k_\perp$ resolution of the PE experiment in the sense of being limited by the physics of the PE process rather than by the measurement accuracy.[12] The averaging over $\delta k_\perp$ can result in *intrinsic shifts* of PE peaks from the energy positions dictated by strict momentum conservation between the initial and final states. Such shifts can occur by different mechanisms, for example:

(i) Sharp asymmetric variations of $M_{fi}$ through the $\delta k_\perp$ interval;

(ii) Averaging of non-linear $E(k_\perp)$ over $\delta k_\perp$ near the band edges, which pushes the PE peaks towards the band interior[12,39] (in other words, the $k_\perp$ broadening results in compression of the dispersion range apparent in the PE spectrum). In our experimental data such in-band shifts are evident, for example, by comparison of the two PE dispersion extremes in the Te 5$p_z^*$ bottom as reached with $E_f\sim$16 eV and ~28 eV: Although they correspond to the *same* initial state in the A point of the BZ, the second extreme appears ~0.1eV higher in $E_i$ due to larger $\delta k_\perp$ value delivered by the final state, see Fig. 5 (*a*). Our PE calculations in comparison with the DT lines in Fig. 8 (*right*) deliver direct estimate of such phenomena: The PE peaks near all band extremes show in-band shifts of the order of 0.2 eV;

Another effect of the $\delta k_\perp$ averaging is formation of dispersionless spectral structures resulting from singularities of the 1-dimensional density of states (1DOS) $\dfrac{\partial k_\perp}{\partial E}$ piling up at the band edges.[11,12] In our experimental data such *1DOS peaks* appear as weak structures visible in the logarithmically scaled -d$^2$I/d$E^2$ map, Fig. 8 (*b, left*), at the Te 5$p_z^*$ band upper edge in a wide $E_f$ region around 15eV. 1DOS peaks can also be identified at the Te 5$p_z^*$ lower edge and Te 5$p_z$ upper edge near $E_f$ of 15eV. As the hole lifetime decrease towards deeper $E_i$ smears the 1DOS singularities at the band edges, these peaks develop only small intensity normally hidden in the slopes of the gross DT peaks, and become visible only in those $E_f$ regions where all DT peaks move to the opposite band edge and their slopes flatten. The experimental 1DOS peaks are well reproduced by the calculations, Fig. 8 (*b, right*).

Beyond the simple averaging picture, intrinsic shifts can also appear due to interference of different Bloch wave constituents of the final state and initial state. For detailed discussion of the interference effects see Ref. 37. Such interference spectral structures deviating from the DT positions occur primarily in the regions where DT branches corresponding to different final bands intersect, for example in the $E_f$ region near 22 eV within the Te $5p_z^*$ band, Fig. 8 (*left*). They are again reproduced by our calculations, Fig. 8 (*right*).

The intrinsic shifts put certain limits on the accuracy of mapping the 3D dispersions with PE spectroscopy. In general, their magnitude scales up with the ratio of $\delta k_\perp$ to the surface-perpendicular BZ extension $k_\perp^{BZ}$.[11,12] These phenomena are therefore of particular concern for layered materials, characterized by small $k_\perp^{BZ}$. In our case, for example, Fig. 5 (*a*) shows that $\delta k_\perp = 2Imk_\perp$ exceeds $0.5|\Gamma A|$, a significant value compared to $k_\perp^{BZ} = 2|\Gamma A|$. One-step PE calculations, delivering reliable estimate of the intrinsic shifts, can be used to account for them in 3D band mapping. The first analysis of the intrinsic shifts in the framework of one-step PE theory in application to layered materials was reported in Ref. 32.

### D. Band mapping

In the band mapping procedure one determines the valence band $E(k_\perp)$ by mapping $E_i$ of the PE peaks against $k_\perp$ determined by $E_f$ of the peaks through the final-state $E(k_\perp)$ dispersion. We have employed here all experimental peaks excluding the regions where the peaks overlapped with each other (and were thus susceptible to the interference effects[37]) as well as the 1DOS peaks. The peak energies were evaluated in $-d^2I/dE^2$. The VLEED derived final states used in the DT plot of Fig. 10 (*b*) were employed. The plot has allowed us to identify each experimental PE peak with one of the band *1-3* in the final state (note that in view of the multitude of all unoccupied bands along $\Gamma A$, see Fig. 5 (*b*), the analysis of $T_\mathbf{k}$ to distinguish the dominant final bands was crucial for such identification). The corresponding $k_\perp$ values were then extracted from the final-state $E(k_\perp)$ dispersions. Note that the use of the true VLEED derived final states takes our band mapping procedure beyond the limitations of the FE-like as well as ground-state DFT approximation for the final states. Neglected however are the slight intrinsic shifts of PE peaks from the DTs positions, see Fig. 8 (*right*).

The results of our band mapping are shown in Fig. 11 superimposed on the theoretical $E(k_\perp)$. Amplitudes of the $-d^2I/dE^2$ peaks, representing amplitude and sharpness of the spectral peaks, are shown in grayscale. The experimental points show very consistent dispersions in both Te $5p_z$ and $5p_z^*$ bands. Note that the band extremes exactly fit the symmetry points $\Gamma$ and A. The remaining minor disparities in the experimental dispersions (apart from the systematic energy shift due to $\Delta\Sigma$ effects in the valence band, see below) are attributed to the intrinsic shifts of the PE peaks. The achieved band mapping consistency should be contrasted to the erratic results for TiTe$_2$ returned by FE-like final bands (see, for example, Ref. 2) as expected from the disagreements in Fig. 10 (*a*). Radical improvements over the FE-like approximation delivered by the use of VLEED derived final bands were also found for VSe$_2$ and TiS$_2$,[7,9] illustrating the importance of non-FE effects in the final states of layered materials, at least for low excitation energies. Furthermore, we have found that the VLEED derived $\Delta\Sigma$ renormalization of the final-state dispersions was vital for the experimental valence band extremes to fit the symmetry points.

### E. Properties of the experimental 3D valence band structure

Consistent control over $k_\perp$ in our band mapping procedure has enabled us to achieve qualitatively new information about the 3D states in the valence band of TiTe$_2$. Experimental $E(k_\perp)$ in Fig. 11 shows the following peculiarities:

(i) At the top of the Te $5p_z^*$ band the dispersion flattens. The experimental points corresponding to the Te $5p_z^*$ band maximum in the $\Gamma$ point appear ~0.35 eV below $E_F$. With the intrinsic shifts here being of the order of 0.1 eV, see Fig. 8 (*b, right*), the $E(k_\perp)$ maximum appears at ~0.25 eV below $E_F$. The appearance of the 1DOS peak coming from the band maximum confirms its position below $E_F$. Therefore, contrary to the early calculations,[2,6] the Te $5p_z^*$ band does not cross $E_F$ and the corresponding FS pocket in the Γ point does not exist. This fact agrees however with the later calculations in Ref. 4 and in this work. Interestingly, the correct position of the Te $5p_z^*$ band is achieved in our calculations only if the spin-orbit interaction is included.

By virtue of the reliable control over $k_\perp$ our analysis gives, to the best of our knowledge, the first experimental evidence of the absence of the Te $5p_z^*$ derived 3D electron pocket in the Γ point. This fact has serious implications for the transport properties of TiTe$_2$, because such a pocket would have delivered a significant isotropic contribution to the electron transport;

(ii) The Te $5p_z^*$ experimental band shows an energy shift relative to the calculated one varying, with the intrinsic shifts deconvoluted, from -0.3 eV at the upper band edge to -0.6 eV at the lower one. For the Te $5p_{xy}$ band the shift is about +0.4 eV. Similarly to the unoccupied states (see II.C.4) such shifts manifest mostly the ΔΣ self-energy corrections to the DFT ground-state band structure. The different and opposite ΔΣ values observed for the Te $5p_z^*$ and $5p_{xy}$ bands identify the band dependence of ΔΣ. Similar differences in ΔΣ between the valence σ- and π-states were observed for graphite.[39]

It should be noted that our calculations and the pseudopotential Gaussian orbital calculations from Ref. 4 both locate the Te $5p_z$ and Te $5p_z^*$ bands by ~0.5eV lower in energy compared to the ASW calculations from Ref. 2, in closer agreement with the experiment. The discrepancy occurs presumably because the accuracy of the ASW calculations was limited by the use of a muffin-tin potentia

## IV. CONCLUSION

We have investigated 3D effects in the electronic structure of the TiTe$_2$ prototype layered material, focusing on the layer-perpendicular band dispersion along the ΓA line. The VLEED experiment, supported by calculations of the complex band structure $E(\text{Re } k_\perp + i\text{Im}k_\perp)$ within the full-potential ELAPW formalism, has been used to determine the dispersions and lifetimes of the final states engaged in the PE process. The PE experimental data, interpreted on the basis of the VLEED derived final states, has yielded consistent experimental $E(k_\perp)$ dispersions in the valence band. The PE experiment was supported by calculations based on a new Bloch waves formalism within the one-step PE theory, which has for the first time delivered accurate description of PE spectra of TiTe$_2$. Of specific results, we have found that:

(i) The final states of TiTe$_2$ show strong non-free-electron effects due to scattering off its highly modulated quasi-2D crystal potential. The final states feature multiband structure, each of the bands showing significantly non-parabolic dispersion;

(ii) Consistent PE band mapping of the valence band $k_\perp$ dispersions in TiTe$_2$ critically depends on taking into account the non-free-electron and self-energy effects in the final states;

(iii) The FS of TiTe$_2$ does not have any Te $5p_z^*$ derived sheet along the ΓA line, which excludes the corresponding isotropic component in the transport properties.

## ACKNOWLEDGEMENTS

We acknowledge the support of Deutsche Forschungsgemeinschaft (CL124/5-2 and Forschergruppe FOR 353) and the EC support for the experiments at LURE within the Access to Research Infrastructure program.

**FIGURES**

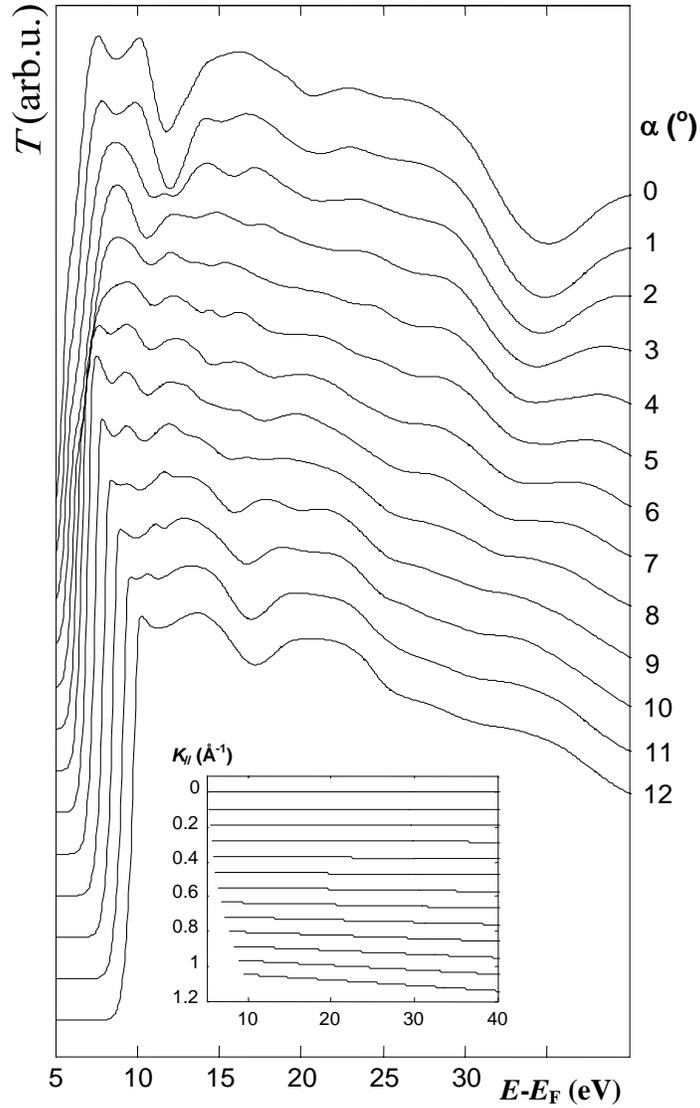

Fig. 1. Experimental VLEED angle-dependent electron transmission spectra $T(E)$ measured along the $\overline{\Gamma K}$ azimuth of the surface BZ at the indicated sample rotation angles. The inset shows the corresponding energy dependences of incidence $\mathbf{K}_{//}$. The spectra show prominent structures, reflecting 3-dimensional $E(\mathbf{k})$ band structure of the PE final states.

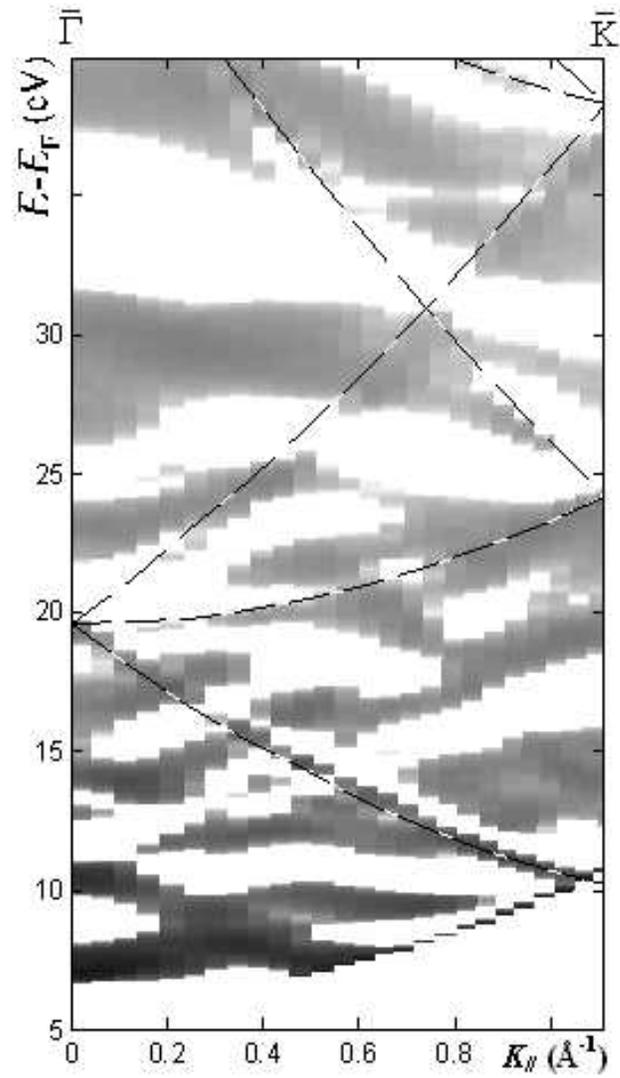

Fig. 2. $\mathbf{K}_{//}$ dispersion map of the experimental spectra from Fig. 1. The shaded and white areas show, respectively, the $T(E)$ maxima and minima energy intervals between the $dT/dE$ extrema reflecting the critical points in 3D final states. Grayscale within the shaded areas characterizes $d^2T/dE^2$ in logarithmic scale. Dashed lines show the diffraction thresholds $E=\hbar^2(\mathbf{K}_{//}+\mathbf{g})^2/2m$ near which the surface resonances are found.

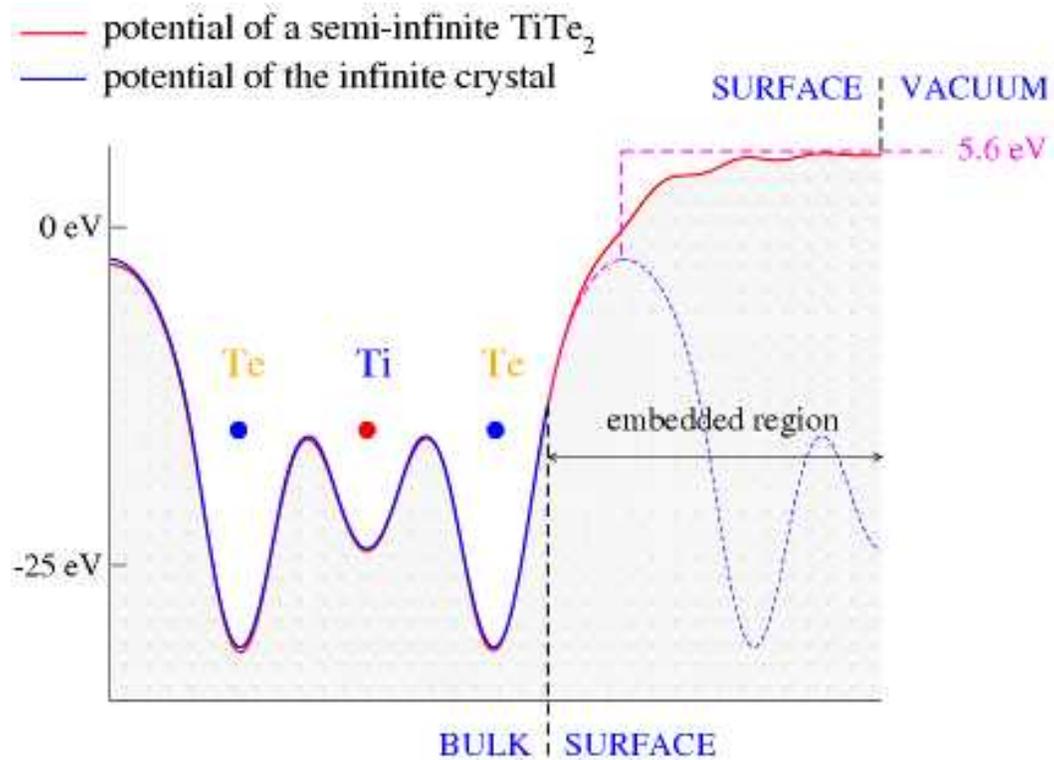

Fig. 3. Calculated potential distribution in the surface region between the bulk crystal and vacuum region. The use of this potential in VLEED calculations considerably improves agreement with experiment over the step-like surface barrier.

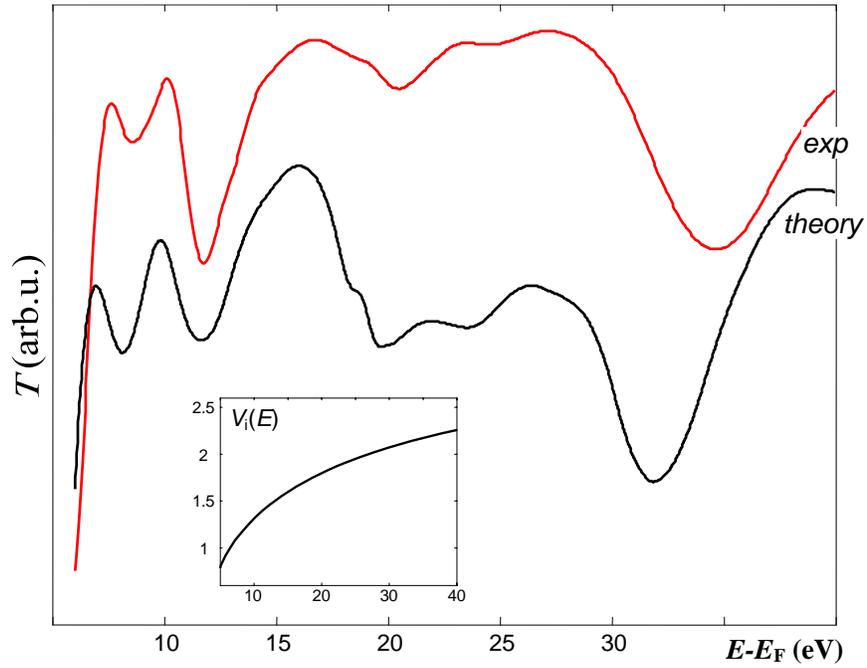

Fig. 4. Theoretical VLEED normal-incidence $T(E)$ spectrum compared with the experimental one from Fig. 1 (linear background subtracted). The inset shows the used $V_i(E)$ dependence. Notable energy shifts between the theoretical and experimental $T(E)$ structures are attributed to the self-energy corrections.

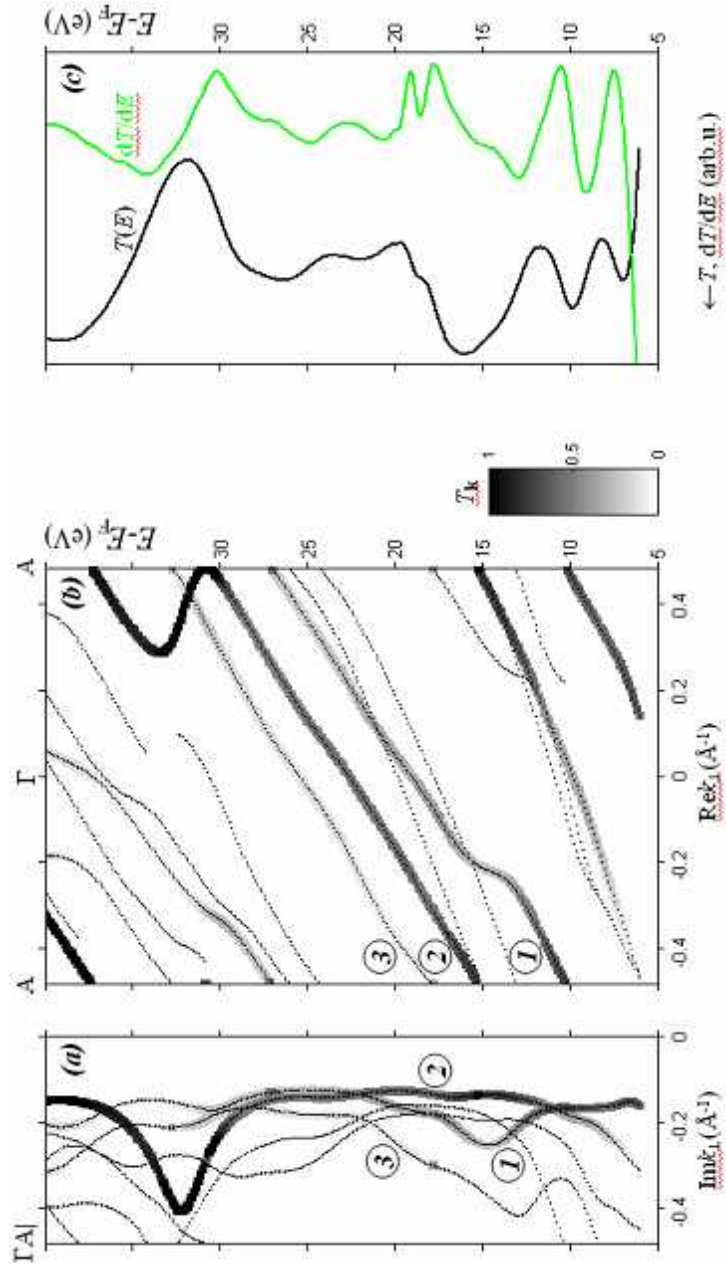

Fig. 5. (*a,b*) Theoretical unoccupied complex $E(\mathbf{k})$ along $\Gamma A$ as a function of $k_\perp = \mathrm{Re}k_\perp + i\mathrm{Im}k_\perp$. Shown are only the Bloch waves characterized by small damping rate $\mathrm{Im}k_\perp < |\Gamma A|$. Grayscale shows the $T_\mathbf{k}$ partial absorbed current contribution of each band to total $T(E)$. Significant $T_\mathbf{k}$ values identify, in the multitude of bands available for given $E$ and $\mathbf{k}_{//}$, the coupling bands dominating in the VLEED and PE processes. Availability of a few such bands (numbered *1* to *3*) and their non-parabolic dispersions are beyond the free-electron-like approximation; (*c*) The corresponding theoretical $T(E)$ and $dT/dE$ spectra.

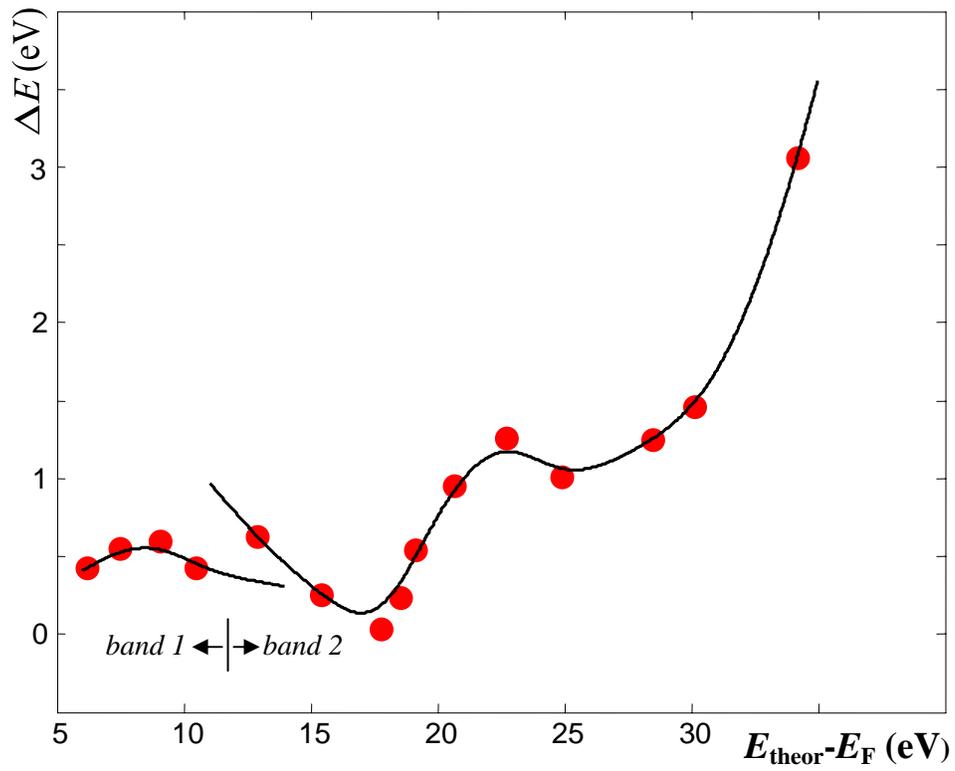

Fig. 6. Energy shifts ΔE between the d$T$/d$E$ extremal points in the experimental and theoretical normal-incidence VLEED spectra as a function of the theoretical d$T$/d$E$ energies. The smooth curves represent correction to the theoretical final bands *1* and *2* used in our PE analysis.

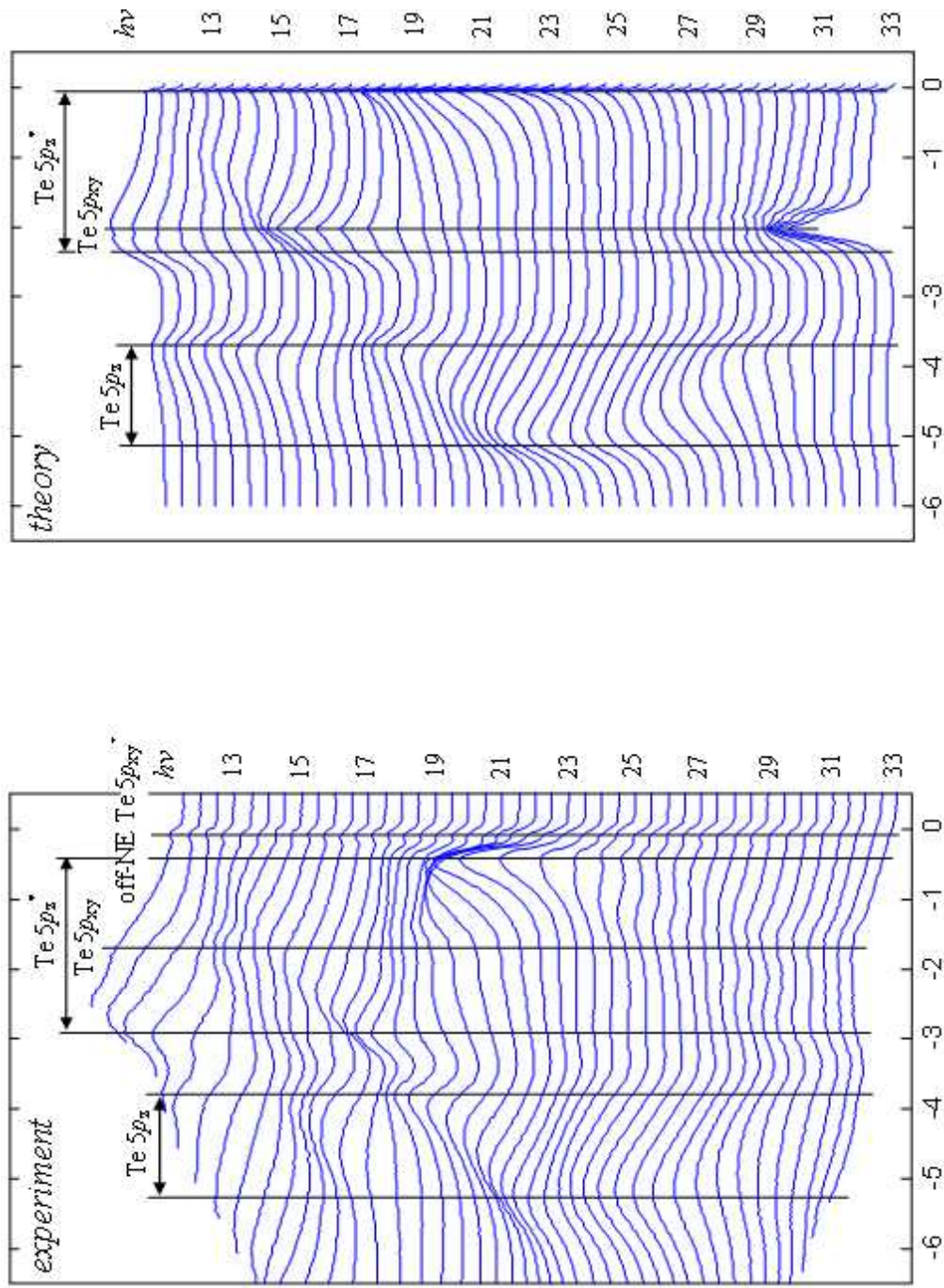

Fig. 7. Experimental (*left panels*) and theoretical (*right*) normal-emission EDC spectra. Dispersion of the spectral peaks with $h\nu$ reflects $E(k_\perp)$ of the indicated valence bands along $\Gamma A$.

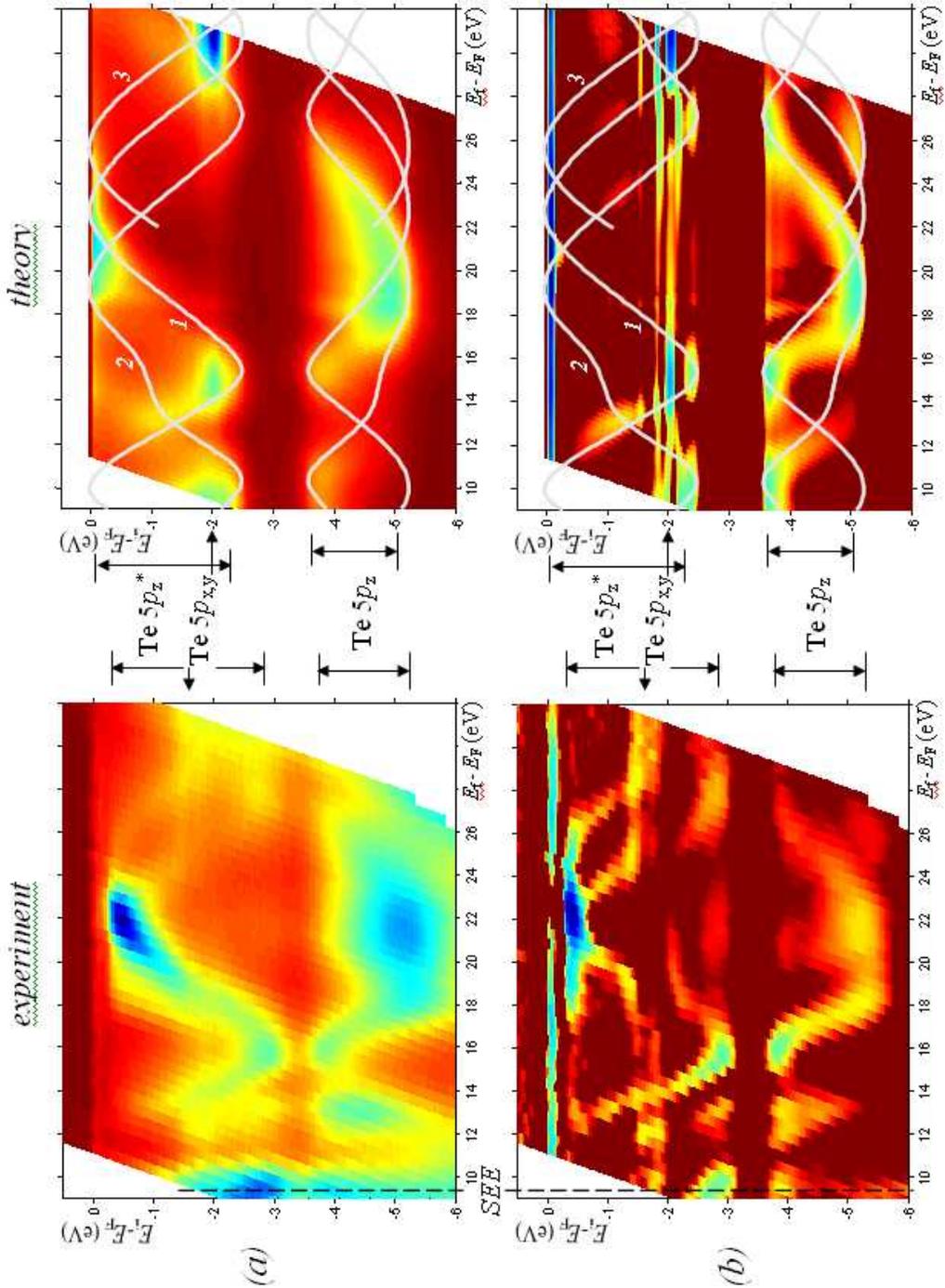

Fig. 8. Experimental (*left panels*) and theoretical (*right*) normal-emission PE data rendered from the EDCs in Fig. 7: (*a*) Intensity map as a function of the initial-state and final-state energies (relative to $E_F$), and (*b*) Map of negative second derivative $-d^2I/dE^2$ of the EDCs (negative values $-d^2I/dE^2 < 0$ set to zero) in a logarithmic colorscale. Multiple dispersion branches within the Te $5p_z$ and Te $5p_z^*$ energy regions demonstrate a multiband composition of the final state, an effect beyond the FE-like approximation. Lines on top of the theoretical maps show the direct transitions (DT) plot constructed with the theoretical initial bands from Fig. 9 and final bands *1-3* having $T_\mathbf{k} > 0.1$ from Fig. 5 (*b*); the DT branches are indexed according to the individual final bands in Fig. 5 (*b*) they originate from. Theoretical PE peaks show notable intrinsic shifts from the DT positions.

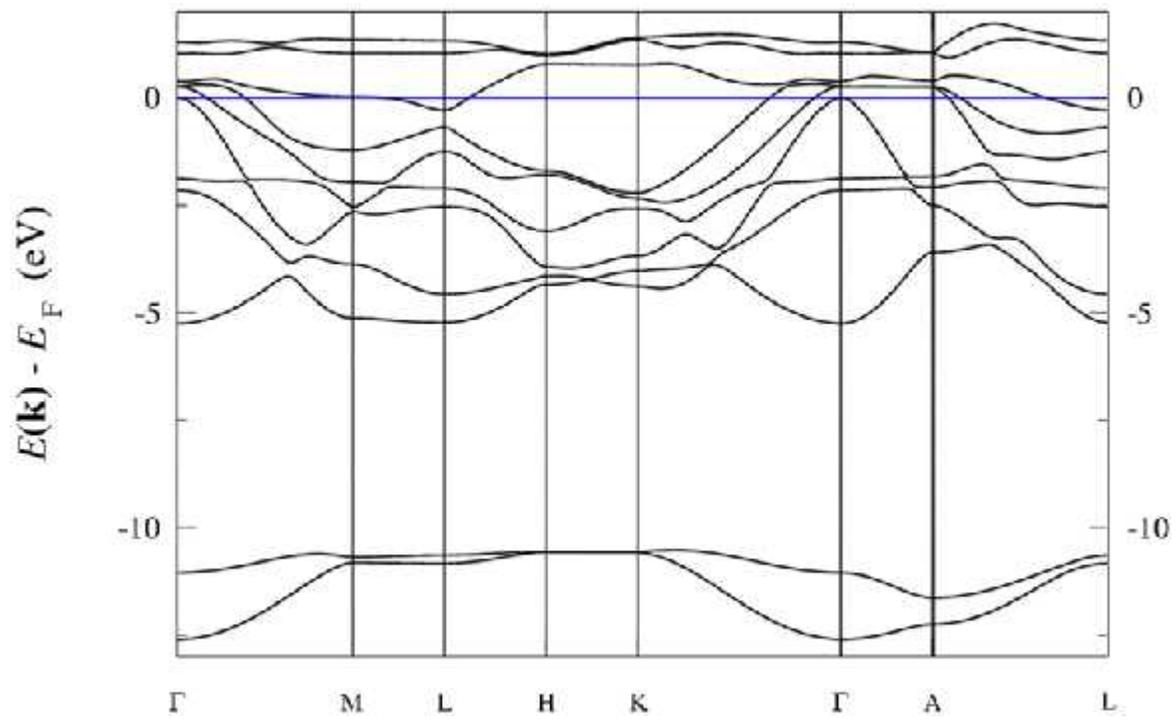

Fig. 9. Theoretical valence band $E(\mathbf{k})$ along representative high-symmetry BZ directions.

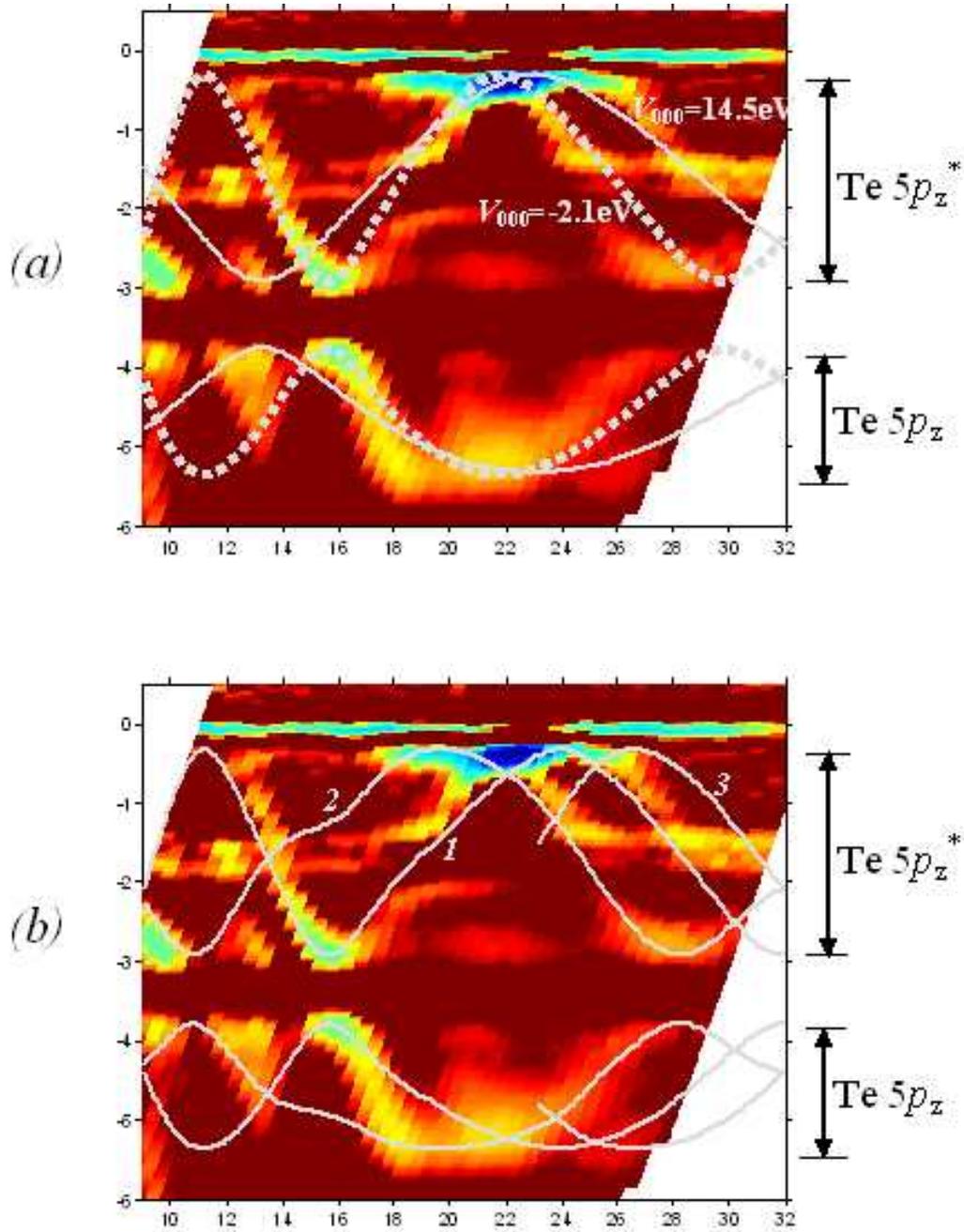

Fig. 10. (*a*) DT plot resulting from the Te $5p_z$ and $5p_z^*$ initial states (the non-dispersive Te $5p_{xy}$ band excluded for clarity) and FE-like final states with conventional $V_{000}$=14.5 eV (*solid lines*) and optimized $V_{000}$=-2.1 eV above the vacuum level (*dashed*), on top of the experimental -$d^2I/dE^2$ data from Fig. 8 (*b*). Limited relevance of this conjecture, in particular failure to reproduce the multiple branches of PE peaks, evidences non-FE effects in the final states; (*b*) DT plot resulting from the same initial states and VLEED derived final states, with the DT branches indexed *1-3* according to the individual final bands with $T_k >$ 0.1 in Fig. 5 (*b*). Due to incorporating the non-FE and self-energy effects in the final states, the plot reproduces most of the peculiarities of the experimental PE data, in particular the multiple dispersion branches.

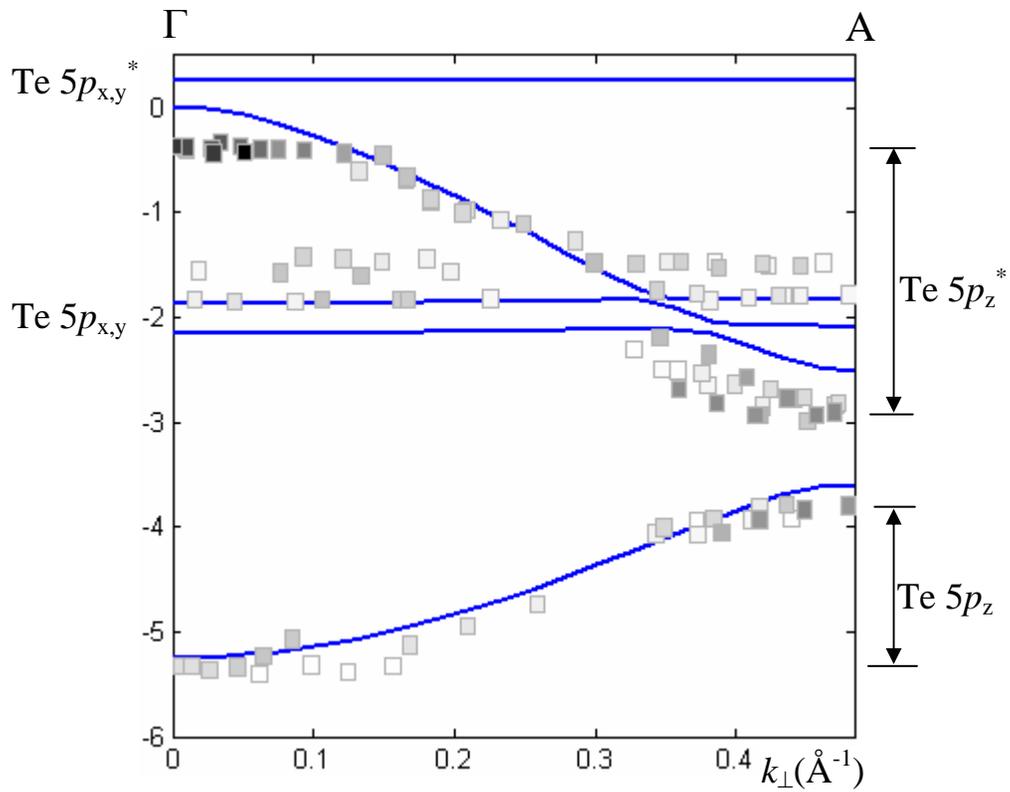

Fig. 11. Experimental valence band $E(k_\perp)$ along ΓA achieved by band mapping with the VLEED derived final states. Amplitude*sharpness of the PE peaks is shown in grayscale (maximum = black). The experimental points show high consistency, contrasting to the results returned by FE-like final bands. The experimental points are superimposed on the LDA-DFT calculation.